\newcounter{myctr}

\documentclass{ws-ijqi}

\begin{document}

\markboth{H.-T. Elze}
{Multipartite cellular automata and the superposition principle}

\catchline{}{}{}{}{}

\title{MULTIPARTITE CELLULAR AUTOMATA AND THE SUPERPOSITION PRINCIPLE  
}

\author{HANS-THOMAS ELZE 
}

\address{Dipartimento di Fisica ``Enrico Fermi'', Universit\`a di Pisa, \\ 
Largo Pontecorvo 3, I-56127 Pisa, Italia
\\ elze@df.unipi.it}



\maketitle

\begin{history}
\received{19 January 2016}
\accepted{22 February 2016}
\end{history}

\begin{abstract}
Cellular automata can show well known features of quantum mechanics, such as a 
linear 
updating rule that resembles a discretized form of the Schr\"odinger equation 
together with its conservation laws. Surprisingly, a whole class of ``natural'' 
Hamiltonian cellular automata, 
which are based entirely on integer-valued variables and couplings and 
derived from an Action Principle, can be mapped reversibly to continuum models with 
the help of Sampling Theory. This results in ``deformed'' quantum mechanical models 
with a finite discreteness scale $l$, which for $l\rightarrow 0$ 
reproduce the familiar continuum limit. Presently, we show, in particular, how such 
automata can form ``multipartite'' systems consistently with the tensor product structures 
of nonrelativistic many-body quantum mechanics, while maintaining the linearity of  
dynamics. Consequently, the Superposition Principle is fully operative already 
on the level of these primordial discrete deterministic automata, including the 
essential quantum effects of interference and entanglement. 
\end{abstract}



\section{Introduction}
The  {\it Cellular Automaton Interpretation} of Quantum Mechanics (QM) has recently 
been proposed by G.~'t\,Hooft.\,\cite{tHooft2014} 
Interest in redesigning the foundations of quantum theory in accordance with essentially classical concepts -- above all, determinism and existence of ontological states 
of reality --  is founded on the observation that quantum 
mechanical features arise in a large variety of deterministic and, in some sense, ``classical'' 
models. {\it E.g.}, the Born rule and collapse of quantum mechanical states  in measurement processes find a surprising and intuitive explanation here, 
where quantum states are superpositions of ontological 
(micro) states, while classical ones are ontological (macro) states, refering to vastly 
different scales in nature.\,\cite{tHooft2014} 

While practically all of these models have been exceptional in that they cannot easily be 
generalized to cover real phenomena, incorporating interactions and relativity, Cellular Automata (CA) may provide the necessary versatility, as we shall presently continue to 
discuss.\,\cite{PRA2014,EmQM13} For an incomplete list of various earlier attempts in 
this field, see, for example,  
Refs.\,\cite{H1,H2,H3,Kleinert,Elze,Groessing,Khrennikov,Margolus,Jizba,Mairi,Isidro,DAriano} 
and  references therein. 

The linearity of quantum mechanics (QM) is a fundamental feature  of unitary dynamics 
embodied in the Schr\"odinger equation. This linearity does not depend on the 
particular object  under study, provided it is sufficiently isolated from 
anything else. It is naturally reflected in the superposition principle and   
entails interference effects and the possibility of non-factorizable states 
of composite objects, {\it i.e.} entanglement in multipartite systems.    

The linearity of QM has been questioned repeatedly and nonlinear modifications 
have been proposed -- not only as suitable approximations for complicated 
many-body dynamics, but especially in order to test experimentally the  
robustness of QM against such {\it nonlinear deformations}. 
This has been thoroughly discussed by T.F. Jordan who presented a `proof 
from within' quantum theory that the theory has to be linear, given the 
essential {\it separability}   
assumption ``... that the system we are considering can be described as
part of a larger system without interaction with the rest of the larger 
system.''\,\cite{Jordan}

Recently, we have considered a seemingly unrelated {\it discrete} 
dynamical theory, {\it i.e.}, which deviates drastically from quantum theory, 
at first sight. 
However, we have shown with the help of Sampling Theory 
that the deterministic mechanics of the class of discrete {\it Hamiltonian CA} can be 
mapped one-to-one to continuum models pertaining to nonrelativistic QM, however, 
modified by the presence of a fundamental time scale.\,\cite{PRA2014,EmQM13} 

For this construction of an intrinsically linear relation between CA 
and QM with a nonzero discreteness scale, 
the consistency of the action principle underlying the discrete dynamics on one side 
and the required locality of the continuum description on the other are compatible only with the linearity of both theories.\,\cite{Discrete14}  
  
The purpose of the present note, in particular, is to study composite objects formed 
from CA subsystems.  -- Clearly, QM is special in that it is characterized not only by interference 
effects, like any classical wave theory would be, but also by the tensor product structures applying for composite systems, which entail the possibility of entanglement. -- 
It is not obvious that CA can form composites which conform with QM, in the limit of negligible discreteness scale. 
This is due to the fact that the state space of Hamiltonian CA is not a complex projective 
space and that the norm of the analogue of state vectors is not conserved by the dynamics; 
instead there is a conserved two-time correlation function, as we shall see, which becomes 
the familiar norm only in the continuum limit.   

In Section\,2., we will briefly review earlier results concerning individual CA which will be useful in the following. One way of composing CA, which is compatible 
with QM, will be shown in Section\,3. Such outside perspective based on CA should eventually lead to additional insight in regard to interference and entanglement.  Concluding remarks are presented 
in Section\,4.  

\section{From action to conservation laws for Hamiltonian CA}
We describe the dynamics of classical Hamiltonian CA with countably many degrees 
of freedom in terms of {\it complex integer-valued} state variables 
$\psi_n^\alpha$, where 
$\alpha\in {\mathbf N_0}$ denote different degrees of freedom and $n\in {\mathbf Z}$  
different states labelled by this discrete {\it clock variable}.  
Various equivalent forms of the action for such CA can conveniently be chosen, as indicated  
earlier.\,\cite{PRA2014} We will employ a particularly compact form 
here, which will be useful in the following, when we  construct composite CA 
in analogy with multipartite QM systems.   
   
Let $\hat H:=\{  H^{\alpha\beta}\}$ denote a self-adjoint complex integer-valued matrix 
that will play the role of the Hamilton operator shortly. Furthermore, we introduce 
the suggestive notation $\dot O_n:=O_{n+1}-O_{n-1}$, for any quantity $O_n$ depending 
on the clock variable $n$. Then, with an 
implicit summation convention for Greek indices, 
$r^\alpha s^\alpha\equiv\sum_\alpha r^\alpha s^\alpha$, we will often simplify the 
notation further by suppressing them altogether, for example, writing 
$\psi_n^{*\alpha}H^{\alpha\beta}\psi_n^\beta\equiv\psi_n^*\hat H\psi_n$. 

Incorporating these conventions, an integer-valued CA action  
${\cal S}$ is defined by:  
\begin{equation}\label{action} 
{\cal S}[\psi ,\psi^*]\; :=\;
\sum_n\big [\frac{1}{2i}(\psi_n^*\dot\psi_n-\dot\psi_n^*\psi_n)+\psi_n^*\hat H\psi_n\big ] 
\;\equiv\;\psi^*\hat{\cal S}\psi 
\;\;, \end{equation} 
with $\psi_n^\alpha$ and $\psi_n^{*\alpha}$ as independent variables; the 
operator $\hat{\cal S}$ will be a useful abbreviation, {\it cf.} Section\,3. 
For the purpose of setting up a variational principle, we introduce    
{\it integer-valued} variations $\delta f$ which are applied to a polynomial $g$ 
as follows: 
\begin{equation}\label{variation} 
\delta_{f}g(f):=[g(f+\delta f)-g(f-\delta f)]/2\delta f 
\;\; , \end{equation} 
and $\delta_fg\equiv 0$, if $\delta f=0$. -- 
We remark that variations  of terms that are 
{\it constant, linear, or quadratic} in integer-valued variables  yield analogous results as  
standard infinitesimal variations of corresponding expressions in the continuum.  --  
Making use of these ingredients, we postulate the variational principle:      
\vskip 0.15cm \noindent 
({\it CA Action Principle}) \hskip 0.15cm   
The discrete evolution of a CA is determined by stationarity of its 
action under arbitrary integer-valued variations of all 
dynamical variables, $\delta {\cal S}=0$.\hfill $\bullet$ \vskip 0.15cm  

It is worth emphasizing several characteristics of this {\it CA Action Principle}:  
\\ \noindent  
{\bf i)} While infinitesimal variations do not conform with integer valuedness, 
there is {\it a priori} no restriction of integer variations. Hence {\it arbitrary}  
integer-valued variations must be admitted. \\ \noindent 
{\bf ii)} One could imagine contributions to the action (\ref{action})  which are of 
higher than second order in $\psi_n$ or $\psi_n^*$. 
However, in view of arbitrary  variations 
$\delta \psi_n^\alpha$ and $\delta \psi_n^{*\alpha}$, 
such additional contributions to the action 
must be absent for consistency.  
Otherwise the number of equations of motion 
generated by variation of the action, according to Eq.\,(\ref{variation}), 
would exceed the number of variables. Yet a  
limited number of such remainder terms, which are nonzero only for some fixed values 
of $n$, could serve to encode  
the {\it initial conditions} for the CA evolution.    

We have shown earlier that these features of the {\it CA Action Principle} 
are essential in constructing a map between Hamiltonian CA and equivalent quantum mechanical continuum models.\,\cite{PRA2014} -- In addition, generalizations of the 
variations defined in Eq.\,(\ref{variation}) have been considered, which allow  
higher than 
second order polynomial terms in the action. However, while leading to consistent 
discrete equations of motion, these equations are beset with undesirable 
nonlocal features in the corresponding continuum model 
description.\,\cite{Discrete14}   

\subsection{The equations of motion} 
It is straightforward now to obtain the equations of motion determined by the 
 {\it CA Action Principle} for the action ${\cal S}$ given by Eq.\,(\ref{action}) with 
the definition of variations provided in Eq.\,(\ref{variation}). Namely, variations  
$\delta\psi_n^*$ and $\delta\psi_n$, respectively, yield 
discrete analogues of the Schr\"odinger equation and its adjoint: 
\begin{eqnarray}\label{delpsistar} 
\dot\psi_n&=&\frac{1}{i}\hat H\psi_n 
\;\;, \\ [1ex] \label{delpsi} 
\dot\psi_n^*&=&-\frac{1}{i}(\hat H\psi_n)^* 
\;\;, \end{eqnarray} 
recalling that $\hat H=\hat H^\dagger$ and $\dot\psi_n =\psi_{n+1}-\psi_{n-1}$, {\it etc.} 
Note that the action ${\cal S}$ vanishes when 
evaluated for solutions of these equations. 

We remark that by setting $\psi_n^\alpha =:x_n^\alpha +ip_n^\alpha$, with real 
integer-valued variables $x_n^\alpha$ and $p_n^\alpha$, and suitably separating real and imaginary parts of Eqs.\,(\ref{delpsistar})--(\ref{delpsi}), the equations 
assume a form that resembles Hamilton's equations for a network of coupled 
discrete classical oscillators:\,\cite{Heslot85,Skinner13}   
\begin{equation}\label{xdotCA} 
\dot x_n^\alpha\;=\;h_S^{\alpha\beta}p_n^\beta +h_A^{\alpha\beta}x_n^\beta  
\;\;,\;  
\;\;\dot p_n^\alpha \;=\;-h_S^{\alpha\beta}x_n^\beta +h_A^{\alpha\beta}p_n^\beta  
\;\;, \end{equation}
where we split the self-adjoint matrix $\hat H$ 
into real integer-valued symmetric and antisymmetric parts, respectively, 
$H^{\alpha\beta}=:h_S^{\alpha\beta}+ih_A^{\alpha\beta}$. -- 
The appearance of these equations has suggested the name 
Hamiltonian CA.\,\cite{PRA2014} 

\subsection{The conservation laws} 
The time-reversal invariant equations of motion that we have obtained 
give rise to conservation laws which are in {\it one-to-one correspondence} with those of 
the related Schr\"odinger equation in the continuum. It is straightforward to verify 
the validity of the following theorem. 

\vskip 0.15cm \noindent 
({\it Theorem\,A}) \hskip 0.15cm For any matrix $\hat G$ that commutes with 
$\hat H$, $[\hat G,\hat H]=0$, there 
is a {\it discrete conservation law}: 
\begin{equation}\label{Gconserv} 
 \psi_n^{\ast\alpha}G^{\alpha\beta}\dot\psi_n^\beta +
\dot\psi_n^{\ast\alpha}G^{\alpha\beta}\psi_n^\beta =0 
\;\;. \end{equation}  
For self-adjoint $\hat G$,  with complex integer elements, 
this relation concerns real integer quantities.\hfill $\bullet$  

By rearranging Eq.\,(\ref{Gconserv}), we can read off the corresponding 
conserved quantity $q_{\hat G}$ (using matrix notation, as before): 
\begin{equation}\label{qG} 
q_{\hat G}:=\psi_n^*\hat G\psi_{n-1}+\psi_{n-1}^*\hat G\psi_n
=\psi_{n+1}^*\hat G\psi_n+\psi_n^*\hat G\psi_{n+1}   
\;\;, \end{equation}  
{\it i.e.} a real integer-valued two-point correlation function which is invariant 
under a shift $n\rightarrow n+m$, $m\in\mathbf{Z}$. -- In particular, for 
$\hat G:=\hat 1$, the corresponding conservation law amounts to a constraint on 
the state variables: 
\begin{equation}\label{normal}  
q_{\hat 1}=2\mbox{Re}\;\psi_n^*\psi_{n-1}
=2\mbox{Re}\;\psi_{n+1}^*\psi_n=\mbox{const}   
\;\;, \end{equation}  
which we anticipate to play a similar role for discrete CA as the familiar normalization 
of state vectors in continuum QM.  
  
For later convenience, we also define the following symmetrized quantity: 
\begin{equation}\label{Q} 
\psi_n^*\hat{\cal Q}\psi_n :=\frac{1}{2}\mbox{Re}\;\psi_n^*(\psi_{n+1}+\psi_{n-1})  
\equiv\frac{1}{2}\mbox{Re}\;\psi_n^{*\alpha}(\psi_{n+1}^\alpha +\psi_{n-1}^\alpha )  
\;\;, \end{equation} 
which, by Eq.\,(\ref{normal}), is conserved as well.   

\subsection{The continuum representation}
Previously we have constructed a one-to-one invertible 
map between the dynamics of discrete Hamiltonian CA and continuum QM in the 
presence of a fundamental time scale.\,\cite{PRA2014,EmQM13,Discrete14} 
Such a finite discreteness scale $l$ implies that continuous time wave functions must be 
{\it bandlimited}, {\it i.e.}, their Fourier transforms have only finite support 
in frequency space, $\omega\in [-\pi /l,\pi /l]$. 
Under these circumstances Sampling Theory  
can be applied, in order to reconstruct continuous time signals, 
wave functions $\psi^\alpha (t)$, from their representative discrete samples, the CA state variables $\psi_n^\alpha$, and {\it vice versa}.\,\cite{Shannon,Jerri,Kempf}

Instead of going through the argument,\,\cite{PRA2014,Discrete14} we  
give the simple mapping rules that result from the reconstruction formula 
provided by {\it Shannon's Theorem}:\,\cite{Shannon,Jerri}   
\begin{eqnarray}\label{psit} 
\psi_n^\alpha &\longmapsto &\;\psi^\alpha (t)
\;\;, \\ [1ex] \label{npm1}
\psi_{n\pm 1}^\alpha &\longmapsto &
\;\exp\big [\mp l\frac{\mbox{d}}{\mbox{d}t}\big ]\psi^\alpha (t)
=\psi^\alpha (t\mp l) 
\;\;, \\ [1ex] \label{samp} 
\psi^\alpha (nl)&\longmapsto &\;\psi_n^\alpha 
\;\;, \end{eqnarray} 
keeping in mind that the continuum wave function is bandlimited. 

With the help of these results, one can map the CA equations of 
motion, in particular Eqs.\,(\ref{delpsistar})--(\ref{delpsi}) to the appropriate 
continuum version. Corresponding to Eqs.\,(\ref{Gconserv})--(\ref{Q}), there exist 
analogous conservation laws and conserved quantities, which can be 
found by applying the mapping rules separately to all wave function factors that appear. 
For example, we obtain from Eq.\,(\ref{Q}) the conserved quantity: 
\begin{eqnarray}\label{Qcont1}  
\mbox{const}=\psi_n^*\hat{\cal Q}\psi_n&\longmapsto&\; 
\psi^*(t)\hat{\cal Q}\psi (t)
=\mbox{Re}\;\psi^*(t)\cosh\big [l\frac{\mbox{d}}{\mbox{d}t}\big ]\psi (t)
\\ [1ex] \label{Qcont2} 
&\;&=
\psi^{*\alpha}(t)\psi^\alpha (t)
+\frac{l^2}{2}\mbox{Re}\;\psi^{*\alpha}(t)\frac{\mbox{d}^2}{\mbox{d}t^2}\psi^\alpha (t) 
+\mbox{O}(l^4)   
\;\;, \end{eqnarray} 
which shows the $l$-dependent corrections to the continuum limit, which here 
amounts to the usual conserved normalization 
$ \psi^{*\alpha}\psi^\alpha =\mbox{const}$\,. 
Similarly, the Schr\"odinger equation and its finite-$l$ corrections are 
obtained.\,\cite{PRA2014}  

This completes our considerations of single Hamiltonian CA, which form the 
basis for the study of multipartite systems.    

\section{Composing multipartite CA}  
Here we address the important question how discrete CA would combine to form 
composite multipartite systems. In particular, two requirements appear naturally, when 
discussing possible constructions. 

Recalling the similarities with QM that we have found, so far, one may wonder whether not only the {\it linearity} of the evolution law  but 
also the {\it tensor product structure} of composite wave functions finds its analogue  here. These are fundamental ingredients of the usual continuum theory, which are reflected in a spectacular manner in interference and entanglement, 
respectively. Which should be recovered, at least, in the continuum limit 
($l\rightarrow 0$) of the CA picture. -- Furthermore, when the discreteness scale $l$ is truly finite, the dynamics of composites of CA which do not interact among each other should lead to {\it no spurious correlations} among them. Such a principle of ``no correlations without interactions'' is  respected more or less explicitly by all known physical theories.\,\cite{Jordan}   
 
We begin by pointing out obstacles which seem to prevent 
satisfying the above requirements, when trying to form composites of Hamiltonian CA.  

The want-to-be 
discrete time derivative introduced before, $\dot O_n:=O_{n+1}-O_{n-1}$, for any quantity $O_n$ depending on the clock variable $n$, which appears all over in the 
CA equations of motion and conservation laws, does not obey the product rule or  
{\it Leibniz's rule}: 
\begin{equation}\label{Leibniz} 
\dot {[A_nB_n]}=
\dot A_n\textstyle{\frac{B_{n+1}+B_{n-1}}{2}}+
\textstyle{\frac{A_{n+1}+A_{n-1}}{2}}\dot B_n 
\neq \dot A_nB_n +A_n\dot B_n 
\;\;. \end{equation} 
Similar observations can be expected for other definitions one might come up with. 
Let us ignore this for a moment and, 
by way analogy with the single-CA Eq.\,(\ref{delpsistar}),  
look at the following multi-CA equation of motion: 
\begin{equation}\label{PSIeq} 
\dot\Psi_n=\frac{1}{i}\hat H_0\Psi_n 
\;\;, \end{equation} 
where $\hat H_0$  may describe a block-diagonal Hamiltonian in the absence of 
interactions among the CA. Then, through Eq.\,(\ref{Leibniz}),  the 
expected {\it factorization} 
of Eq.\,(\ref{PSIeq}) is hindered on the left-hand side, since unphysical correlations  
will be produced among the components of a factorized wave function, such as   
\begin{equation}\label{PSI} 
\Psi_n^{\alpha\beta\gamma\cdots}=
\psi_n^\alpha\phi_n^\beta\kappa_n^\gamma\cdots 
\;\;, \end{equation}  
and, correspondingly, for a superposition of such factorized terms. 
Thus, for a bipartite system we have:  
$\dot \Psi_n^{\alpha\beta}=
\dot\psi_n^\alpha (\phi_{n+1}^\beta+\phi_{n-1}^\beta )/2+ 
\psi\leftrightarrow\phi 
\neq\dot\psi_n^\alpha\phi_n^\beta +\psi_n^\alpha\dot\phi_n^\beta$. 

Furthermore, applying the mapping rules of Section\,2.3, before taking the 
limit $l\rightarrow 0$, we find that the bilinear terms here do not 
converge to the appropriate QM expression.  
Of course, it should be   
$\partial_t(\psi^\alpha\phi^\beta )=(\partial_t\psi^\alpha )\phi^\beta +\psi^\alpha \partial_t\phi^\beta$, in order to allow   
the decoupling of two subsystems that do not interact.  

However, this latter problem is a general one of nonlinear terms in the equations 
of motion of discrete CA, 
which we discussed before:\,\cite{Discrete14}  {\it The linear map provided by 
Shannon's Theorem does not commute with the multiplication implied by the 
nonlinearities.}  In particular, the map of a bilinear term is not equal to the bilinear 
term of its mapped entries, symbolically: $$A_nB_n\equiv C_n\mapsto C(t)\;\neq\; A(t)B(t)\;\;,$$ 
where $A_n\mapsto A(t)$ and $B_n\mapsto B(t)$, as follows from the 
explicit reconstruction formula (or any variant thereof that is linear).\,\cite{PRA2014,Shannon,Jerri} 
In fact, this problem arises also on the right-hand side of Eq.\,(\ref{PSIeq}), 
when trying to map a factorized wave function to its continuous time description.     

\subsection{The many-time formulation} 
It appears that 
the  difficulties arise from the implicit assumption that the components of a multipartite CA are {\it synchronized} to the extent that they share a common clock 
variable $n$. We consider a radical way out of the impasse encountered 
by resorting to the {\it many-time} formalism, which means giving up synchronization 
among parts of the composite CA by introducing a set of clock variables, 
$\{ n(1),\;\dots ,\;n(m)\}$, one for each one out of $m$ components.  

This may come as a surprise in the present nonrelativistic context, since the 
many-time formalism has been introduced by Dirac, Tomonaga, and Schwinger in 
their respective formulations of relativistically covariant many-particle 
QM or quantum field theory, where a global synchronization cannot be 
maintained.\,\cite{Dirac,Tomonaga,Schwinger}  
  
Replacing the single-CA action of Eq.\,(\ref{action}), we define here
the integer-valued multipartite-CA action by: 
\begin{equation}\label{maction} 
{\cal S}[\Psi ,\Psi^*] :=\Psi^*\big (
\sum_{k=1}^m\hat{\cal S}_{(k)}\;+\;\hat{\cal I}\big )\Psi
\;\;, \end{equation} 
with $\Psi :=\Psi^{\alpha_1\dots\alpha_m}_{n_1\dots n_m}$ and, correspondingly,  $\Psi^*$ as independent {\it complex integer-valued} variables; 
the self-adjoint operator $\hat{\cal I}$ incorporates interactions between different CA; 
whereas
$\hat{\cal S}_{(k)}$ is as introduced in Eq.\,(\ref{action}), with the subscript 
$_{(k)}$ indicating that it acts {\it exclusively} on the pair of indices pertaining to 
the $k$-th single-CA subsystem:   
\begin{equation}\label{Sopk} 
\Psi^*\hat{\cal S}_{(k)}\Psi:=
\sum_{\{ n_k\} }\big[ (\mbox{Im}\;
\Psi_{\dots n_{k}\dots}^{*\dots\alpha_{k}\dots}
\;\dot\Psi_{\dots n_k\dots}^{\dots\alpha_k\dots}
\;+\;\Psi_{\dots n_{k}\dots}^{*\dots\alpha_{k},\dots}\;
H_{(k)}^{\alpha_k\beta_k}\Psi_{\dots n_k\dots}^{\dots\beta_k\dots}
\big ]
\;\;, \end{equation} 
with summation over {\it all} clock variables (summation over 
twice appearing Greek 
indices remains understood); the $\dot{\phantom .}$-operation, however, acts only with 
respect to the explicitly indicated $n_k$, $\dot f(n_k):=f(n_k+1)-f(n_k-1)$, while the 
single-CA Hamiltonian, $\hat H_{(k)}$, requires a matrix multiplication, as before.  

Obviously, we can apply the {\it CA Action Principle} of Section\,2. to the present situation as well, with the generalized action of Eq.\,(\ref{maction}), in particular. This results in the following discrete equations of motion: 
\begin{equation}\label{mEoM} 
\sum_{k=1}^m\dot\Psi_{\dots n_k\dots}^{\dots\alpha_k\dots}
\;=\;\frac{1}{i}\big (\sum_{k=1}^mH_{(k)}^{\alpha_k\beta_k}
\Psi_{\dots n_k\dots}^{\dots\beta_k\dots} 
\;+\;{\cal I}^{\dots\alpha_k\dots\;\beta_1\dots\beta_m}
\Psi_{\dots n_k\dots}^{\beta_1\dots\beta_m} 
\big )
\;\;, \end{equation}  
together with the adjoint equations; here the interaction $\hat{\cal I}$, 
like $\hat H_{(k)}$, is assumed to be independent of the clock variables and    
the $\dot{\phantom .}$-operation acts only with respect to 
$n_k$ in the $k$-th term on the left-hand side.   

Let us verify that the many-time formulation presented here avoids  
the problems of a single-time multi-CA equation, such as Eq.\,(\ref{PSIeq}), which we 
pointed out. 

First of all, in the absence of interactions  
between CA subsystems, $\hat{\cal I}\equiv 0$, it is {\it sufficient} for a solution of  
Eqs.\,(\ref{mEoM}) that the multi-CA wave function factorizes:  
\begin{equation}\label{PSIfact} 
\Psi =\prod_{k=1}^m\psi_{n_k}^{\alpha_k} 
\;\;, \end{equation} 
which differs from Eq.\,(\ref{PSI}) by the presence of an individual clock variable 
for each  component CA,  $\{ n_k,\;k=1,\dots ,m\}$, or is a superposition of 
such factorized wave functions, and that each factor solves the 
appropriate single-CA equation of motion (as before, cf. Section\,2.1.): 
\begin{equation}\label{sEoM} 
\dot\psi_{n_k}^{\alpha_k}=\frac{1}{i}H_{(k)}^{\alpha_k\beta_k}
\psi_{n_k}^{\beta_k}\;\;,\;\;\;k=1,\dots ,m
\;\;. \end{equation}  
Thus, {\it no unphysical correlations} are introduced among independent CA subsystems 
which do not interact with each other.  

Secondly, the 
{\it continuous multi-time equations} corresponding to Eqs.\,(\ref{mEoM}) are obtained 
by applying the mapping rules given in Section\,2.3. to the discrete equations, as 
determined by Sampling Theory. Presently, there arises no problem of incompatibility between multiplication according to nonlinear terms {\it vs.} linear mapping  according to 
{\it Shannon's Theorem}, since a separate mapping has to be applied for each one of the 
individual clock variables. This effectively replaces $n_k\rightarrow t_k,\;k=1,\dots,m$, 
where $t_k$ is a continuous real time variable.  In this way, the following {\it modified 
multi-time Schr\"odinger equation} is obtained:  
\begin{equation}\label{mSchroed} 
\sum_{k=1}^m\sinh\big [l\frac{\mbox{d}}{\mbox{d}t_k}\big ] \Psi_{\dots t_k\dots}^{\dots\alpha_k\dots}
\;=\;\frac{1}{i}\big (\sum_{k=1}^mH_{(k)}^{\alpha_k\beta_k}
\Psi_{\dots t_k\dots}^{\dots\beta_k\dots} 
\;+\;{\cal I}^{\dots\alpha_k\dots\;\beta_1\dots\beta_m}
\Psi_{\dots t_k\dots}^{\beta_1\dots\beta_m} 
\big )
\;\;, \end{equation}  
where an overall factor of two from the left-hand side has been absorbed into the matrices  on the right.  Note that the wave function $\Psi$ is bandlimited, by construction,  
with respect to each variable $t_k$. 

Performing the continuum limit, $l\rightarrow 0$, we arrive at the  
multi-time Schr\"odinger equation (one power of $l^{-1}$ providing the 
physical dimension of $\hat H_{(k)}$ and $\hat {\cal I}$) considered by Dirac and 
Tomonaga.\,\cite{Dirac,Tomonaga} However, when $l$ is fixed and finite, modifications 
in the form of powers of $l\mbox{d}/\mbox{d}t_k$ arise on its left-hand side.     

Furthermore, in the present nonrelativistic context, it may be appropriate to identify  
$t_k\equiv t,\;k=1,...,m$, in which case the operator on the left-hand side of 
Eq.\,(\ref{mSchroed}), for $l\rightarrow 0$, can be simply replaced by 
$\mbox{d}/\mbox{d}t$, which results in the usual (single-time) 
{\it many-body Schr\"odinger equation}.  

\subsubsection{The conservation laws of multipartite CA} 
Symbolically, the equivalent many-time equations (\ref{mEoM}) and (\ref{mSchroed}) 
are obviously both of the form:  
\begin{equation}\label{msymb} 
\hat {\cal D}\Psi =\frac{1}{i}(\hat {\cal H}+\hat {\cal I})\Psi 
\;\;, \end{equation} 
to be complemented by corresponding adjoint equations. Then, for any operator 
$\hat {\cal G}$, such that $[ \hat {\cal G},\hat {\cal H}+\hat {\cal I}]=0$, we find 
immediately the generalization of {\it Theorem\,A} of Section\,2.2., 
namely the {\it discrete conservation law for multipartite CA}: 
\begin{equation}\label{mTheoremA} 
\Psi^*\hat {\cal G}\hat {\cal D}\Psi +(\hat {\cal D}\Psi^*)\hat {\cal G}\Psi =0 
\;\;, \end{equation}  
valid for the discrete {\it and} continuous time descriptions with the obvious 
explicit form of ${\cal D}\Psi^{(*)}$ inserted, respectively, according to the left-hand 
sides of Eqs.\,(\ref{mEoM}) and (\ref{mSchroed}). 

This, in turn, leads to conserved quantities, to be compared with 
Eqs.\,(\ref{qG})--(\ref{normal}) before. Here we are particularly interested in the case 
$\hat {\cal G}:=\hat 1$, which yields as conserved quantity: 
\begin{eqnarray}\label{mQ} 
\Psi^*\hat{\cal Q}\Psi &:=&  
\sum_{k=1}^m\Psi^{*\alpha_1\dots\alpha_m}_{n_1\dots n_m}\hat{\cal Q}_{(k)}
\Psi^{\alpha_1\dots\alpha_m}_{n_1\dots n_m}  
\\ [1ex] \label{mQc} 
&=&\mbox{Re}\sum_{k=1}^m\Psi^{*\alpha_1\dots\alpha_m}_{t_1\dots t_m} 
\cosh\big [l\frac{\mbox{d}}{\mbox{d}t_k}\big ]\Psi^{\alpha_1\dots\alpha_m}_{t_1\dots t_m}
\\ [1ex]\label{mQcl}   
&\stackrel{l\rightarrow 0}{\longrightarrow}&\;
m\cdot\Psi^{*\alpha_1\dots\alpha_m}_{t_1\dots t_m}
\Psi^{\alpha_1\dots\alpha_m}_{t_1\dots t_m}= m\cdot
|\Psi_{t_1\dots t_m}|^2  
\;\;, \end{eqnarray} 
where subscript $\phantom ._{(k)}$ serves to indicate on which one of the discrete 
clock variables, namely $n_k$, the operator $\hat{\cal Q}$ acts, which has been introduced in Eq.\,(\ref{Q}); the second and third equalities, respectively, present the corresponding 
continuous multi-time quantity and its continuum limit, cf. 
Eqs.\,(\ref{Qcont1})--(\ref{Qcont2}).  This is the 
{\it wave function normalization} in the multi-time formulation\;\cite{Tomonaga};  
when it is appropriate to identify  
$t_k\equiv t,\;k=1,...,m$, the usual many-body wave function normalization follows.  
Included here is, of course, also the case of a factorized wave function as in 
Eq.\,(\ref{PSIfact}).  

\subsubsection{The Superposition Principle in composite Hamiltonian CA}   
The equivalent discrete or continuous many-time equations (\ref{mEoM}) and 
(\ref{mSchroed}) are both linear in the CA wave function $\Psi$. Therefore, superpositions 
of solutions of these equations also present solutions. Thus, the {\it Superposition 
Principle} holds  not only for single but for multipartite Hamiltonian CA as well.    

As in the case of single CA, this entails the fact that already these discrete systems 
 -- with all variables, parameters, {\it etc.} being (complex) integer-valued -- can 
produce {\it interference} effects as in quantum mechanics. Even more interesting, their 
composites can also show {\it entanglement}, which is deemed an essential feature of QM.    
This follows from the form of the equations of motion, which allow for superpositions 
of factorized states, {\it cf.} Eq.\,(\ref{PSIfact}). 

For example, in the bipartite case 
($k=1,2$), assuming that the individual CA are characterized by two degrees of freedom 
($\alpha_k=0,1$), a time dependent analogue of one of the the well known 
{\it Bell states}, the totally antisymmetric one, is given by: 
\begin{equation}\label{Bell} 
\Psi\propto\psi_{n_1}^{\alpha_1=0}\psi_{n_2}^{\alpha_2=1}
-\psi_{n_1}^{\alpha_1=1}\psi_{n_2}^{\alpha_2=0} 
\;\;, \end{equation} 
which may be a solution of appropriate discrete equations of motion. 
  
However, a word of warning is in order here. We have freely used expressions familiar 
in QM, such as ``wave functions'' and ``states'', in particular. These are usually taken to 
invoke the notion of vectors in a {\it Hilbert space}, which becomes a complex projective space upon normalization of the vectors. 

As we have seen already in Section\,2.2., 
see Eqs.\,(\ref{normal})--(\ref{Q}), or Section\,3.1.1., see Eqs.\,(\ref{mQ})--(\ref{mQcl}), as 
long as the CA are truly discrete ($l\neq 0$), the normalization (squared) of vectors is not among the conserved quantities, hence not applicable, but is replaced by a conserved 
(many-)time correlation function instead. 

Furthermore, despite 
close resemblance, the envisaged space of states strictly speaking is not a 
Hilbert space, since it fails in two respects: the vector-space and completeness 
properties are missing.  --  First of all, 
the relevant {\it Gaussian integers} (complex integer-valued numbers) are not 
complete. Hence the completeness property of the space of states is lacking, which is 
built here with these integers as underlying scalars. Secondly, the integer numbers {\it only} 
featuring in all aspects of the CA do not form a 
{\it field} but a {\it commutative ring} (for the multiplication of vectors by such scalars 
there is no multiplicative inverse, such as exists, {\it e.g.}, for rational,  
real, or complex numbers). Therefore, we cannot form a vector space over a field, as 
usual in QM, 
but have to replace it  by a more general structure. This is known as a 
module over a  ring, in the present case a {\it module over the commutative ring of 
Gaussian integers}. It allows the construction of a linear space endowed with an 
integer-valued scalar product, {\it i.e.} a {\it unitary space}. Taking its incompleteness 
into account, then, the space of states in the presented CA theory can be classified as a 
{\it pre-Hilbert module over the commutative ring of Gaussian integers}.\,\footnote{We 
thank a referee 
for his constructive criticism of our earlier presentation of this point, helping to clarify 
the issues involved.}

We conclude that superpositions of states, interference effects, and entanglement, as 
in quantum mechanics, all find their correspondents already on the ``primitive'' level 
of the presently considered natural Hamiltonian CA, discrete single or multipartite systems which are characterized by (complex) integer-valued variables and couplings.    

\section{Conclusion} 
We have presented a brief review of earlier work which has demonstrated surprising quantum features arising in integer-valued, hence ``natural'', {\it Hamiltonian cellular 
automata}.\,\cite{PRA2014,EmQM13,Discrete14,WignerSymp13}  The study of this particular class of CA is motivated by 't\,Hooft's {\it Cellular Automaton 
Interpretation of QM} 
elaborated in Ref.\,\cite{tHooft2014} and various recent attempts to construct models 
which serve to illustrate indeed that QM (or, at least, essential features thereof) can 
be understood to emerge from pre-quantum deterministic dynamics beneath.    

The single CA we considered previously allowed practically for the first time to 
reconstruct quantum mechanical models  with nontrivial Hamiltonians in terms of 
such deterministic systems with a finite discreteness scale. 

Presently, we have  extended 
this study by describing {\it multipartite systems}, analogous to many-body QM. 
Not only is this useful for the construction of more complex models {\it per se} (with 
a richer structure of energy spectra, in particular), but it is also necessary, in order to 
research the equivalent of the {\it Superposition Principle} of QM, if any, on the CA level. 
Thus, we find that it can be introduced already there to the fullest extent, 
compatible with a tensor product structure of multipartite states, 
which entails not only the possibility of their {\it interference} but also of their 
{\it entanglement}. 

Surprisingly, we have been forced -- in our approach employing Sampling Theory for the map between CA and an equivalent continuum picture -- to introduce a 
many-time formulation, which only appeared in relativistic quantum mechanics before, in 
the way introduced by Dirac, Tomonaga, and Schwinger.\,\cite{Dirac,Tomonaga,Schwinger} 
This may point towards a crucial further step in these developments, which is still missing, 
namely a CA model of {\it interacting quantum fields}. It is hard to envisage 
such a picture of dynamical fields spread out in spacetime without the possibility of  
multipartite CA with quantumlike features, which we have presently constructed. 
Yet further conceptual advances seem necessary, in order to  
arrive at a relativistic quantum field theory departing from pre-quantum 
cellular automata. 

\section*{Acknowledgments}
It is a pleasure to thank A.\,Khrennikov for inviting me to the conference ``Quantum Theory: from foundations to technologies'' (Vaxjoe, June 2015), where part of this work was presented, N.\,Margolus for correspondence, and G.\,'t~Hooft for discussions.   


\end{document}